\documentclass[aps,tightenlines,prc,preprint,showpacs,showkeys,floatfix]{revtex4}
\usepackage{epsfig}
\newcommand{\bfDelta}{\mbox{\boldmath $\Delta$}}
\newcommand{\boldsigma}{\mbox{\boldmath $\sigma$}}

\def\poinc{Poincar\'{e} }

\def\bfq {{\bf q}}

\def\bfK{{\bf K}}

\def\bfk{{\bf k}}
\def\bfp{{\bf p}}

\def\be{\begin{equation}}
 \def \ee{\end{equation}}
\def\bea{\begin{eqnarray}}
  \def\eea{\end{eqnarray}}
\newcommand{\eqn} {Eq.~(\ref )}

\def\notK{{\not\! K}}

\def\noteps{{\not\!\epsilon}}
\def\notepsp{{\not\!{\epsilon\;'}^*}}

\def\poinc{Poincar\'{e} }
\def\bfq {{\bf q}}
\def\bfK{{\bf K}}

\def\bfk{{\bf k}}
\def\bfp{{\bf p}}  
\def\be{\begin{equation}}
 \def \ee{\end{equation}}
\def\bea{\begin{eqnarray}}
  \def\eea{\end{eqnarray}}
\def\eqn {Eq.~(\ref }
\newcommand{\bmath}[1]{\mbox{\boldmath${#1}$}}

\newcommand{\beps}{\bmath{\epsilon}}

\date{\today}
\begin{document}
\vskip1in\vskip1in\vskip1in\vskip1in
\preprint{NT@UW-04-002,ECT-03-034,LBNL-54622}
\title{Handling the Handbag Diagram in Compton Scattering on the Proton}

\author{Gerald A. Miller}
\affiliation{Department of Physics, University of Washington, Seattle, 
WA 98195-1560}

\begin{abstract}
\poinc invariance,
gauge invariance, conservation of parity and time reversal invariance
are  respected in an impulse approximation evaluation of the
handbag diagram. Proton wave functions, previously constrained
by comparison with  measured form factors, that incorporate the influence
of quark transverse and 
orbital angular momentum (and the corresponding violation
of proton helicity conservation) are used.  Computed cross sections
are found to be 
in reasonably good agreement with early measurements.
The helicity correlation between the incident photon and outgoing proton,
 $K_{LL}$, is both large and positive at back angles.
 For
 photon laboratory energies of $\le$ 6 GeV,
 we find that $K_{LL}\ne A_{LL}$, and
 $D_{LL}\ne1$.
\end{abstract}

\pacs{13.60.Fz} 
\keywords{Compton Scattering, high momentum transfer, relativistic constituent
quark models}

\maketitle

The recent and planned experimental accessibility\cite{nathan,bogdan} of
 Real Compton Scattering  on the proton  at large momentum
transfer
make this reaction a promising new probe of  
short distance structure. The scattering amplitude  
 depends on the square of the charge of a struck 
quark, and so provides a ground-state to ground-state matrix element
that is different than the ones involved in the electromagnetic form
factors\cite{Jones:1999rz}.

\begin{figure}
\begin{center}
        \leavevmode
        \epsfxsize=0.75\textwidth
        \epsfbox{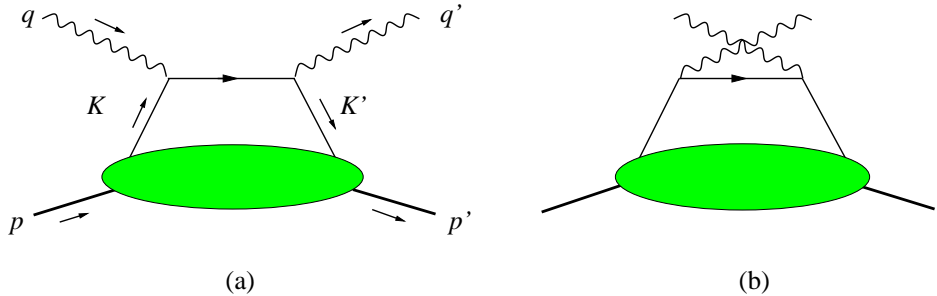}
\end{center}
\caption{\label{fig:hb}(Color Online)  Direct (a) and crossed (b) handbag 
 graphs for the Compton amplitude.}
\end{figure}

One 
 primary goal  has been to  determine the dominant reaction mechanism
that allows the proton to   accommodate  the large momentum transfer while
remaining a proton.
According to perturbative QCD (pQCD)\cite{Kronfeld,Farrar} the
 three active valence quarks  share the momentum transfer via
the exchange of two gluons that each carry a large momentum.
In the pQCD treatment, hard gluon exchanges are 
included in an effective current operator, and so can be taken to 
occur within the time
duration of the reaction.  The most recent calculations
\cite{Brooks:2000nb} 
find
that this mechanism yields cross sections
that are about 10 times smaller than  ones measured at photon energies  of 
6 GeV or less.  
 Another approach
uses  overlaps of soft non-perturbative wave functions\cite{Radyushkin1,Diehl}.
As noted in the review\cite{Diehl:2003ny}, 
the invariant amplitude is  obtained by evaluating
the so-called handbag diagrams  of Fig.~\ref{fig:hb}. The large momentum
transfer occurs on a single quark, with a probability amplitude determined
by the overlap of the initial and final state wave functions. 
Here                                      the high momentum
transfer is accommodated by the exchange of an uncountable number of  gluons
that occurs either before and  after the reaction takes place. 
Such calculations have had reasonable success in reproducing the measurements
\cite{Diehl,hkm}.

Previous treatments of the handbag diagram\cite{Radyushkin1,Diehl}
have provided a  unifying relation between elastic form factors, real Compton
scattering and virtual Compton scattering. 
This approach has led to 
 interesting predictions that are testable 
experimentally. Our concern here is with understanding the limitations
of the approximations.
One  involves asserting that the longitudinal momentum
of the proton is carried by  a single quark\cite{Diehl}. Another 
involves
the neglect of  the effects of hadron helicity 
flip\cite{Radyushkin1,Diehl} in high momentum transfer exclusive
reactions.
A specific consequence 
 is that the observables $K_{LL}$ (which involves the 
 helicity of the final proton) and $A_{LL}$ (which 
involves the helicity of the initial proton) are predicted to be the
same. Ref.~\cite{hkm} includes corrections
that allow photon and proton helicity flip.

Our purpose here is to
 use a model wave 
function that provides a reasonably good description of all four nucleon
electromagnetic form factors\cite{Frank:1995pv,
Miller:2002qb,
Miller:2002ig} to evaluate the graphs of Fig.~\ref{fig:hb} in a manner that
avoids 
 neglecting the effects 
 of hadronic helicity non-conservation. The essential
feature 
is that relativistic and quark mass effects induce 
significant  quark transverse and orbital angular
momentum that cause violations
of hadronic helicity conservation.

We begin   describing  the  formalism  by  reviewing
 the salient features of the wave functions of  Refs.~\cite{Frank:1995pv,
Miller:2002qb,Miller:2002ig}. 
This model starts with a  wave function for 
3 relativistic constituent quarks:
  \bea
\Psi(p_i)&=&u(p_1) u(p_2) u(p_3)\psi(p_1,p_2,p_3),\label{psieq}\eea
where  $p_i$ represents space, spin and isospin indices:
$p_i=\bfp_i  s_i,\tau_i$ and repeated indices are
summed over.
The spinors $u$ are canonical Dirac spinors.  The components of momenta
are expressed in terms of   light cone  notation: 
 ${\bf p}_i \equiv
(p^+,{\bf p}_\perp)_i$, with
$p^-_i=({p_\perp}_i^2+m^2)/p_i^+$ .
The three momenta $\bf{p}_i  $ of the quarks
can be transformed to the total and relative momenta to facilitate
the separation of the center of mass motion
 as $
\bf {P}= \sum_i{\bf p}_i, 
\xi={p_1^+/(p_1^++p_2^+)},
\eta={p_1^++p_2^+/ P^+},
{\bf k}_\perp =(1-\xi){\bf p}_{1\perp}-\xi {\bf p}_{2\perp}\;,
{\bf K}_\perp =(1-\eta)({\bf p}_{1\perp}
+{\bf p}_{2\perp})-\eta {\bf p}_{3\perp}\;. 
$
One may express the proton wave function in the center of mass  frame in which 
the individual momenta are given by
$ {\bf p}_{1\perp}={\bf k}_\perp+\xi {\bf K}_\perp,\;
{\bf p}_{2\perp}=-{\bf k}_\perp+(1-\xi){\bf K}_\perp\;, 
{\bf p}_{3\perp}=-{\bf K}_\perp.
$
The structure of the color-spin-isospin wave function can be understood in
a familiar form. This eigenstate of spin\cite{bere76,chun91}
is  a product
of an anti-symmetric color wave function  with a 
symmetric flavor-spin-momentum 
wave function, given by  
$\Psi={1\over\sqrt{2}}
\left(\phi_\rho\;\chi_\rho+\phi_\lambda\;\chi_\lambda\right)\Phi, $
  where $\phi_{\rho}$ represents a mixed-antisymmetric and $\phi_{\lambda}$
a   mixed-symmetric flavor wave function and, $\chi_{\rho,\lambda}$
  represents mixed symmetric or anti-symmetric
 spin wave functions in terms of Dirac  spinors. The lower components of these
contain terms in which the spin of the quark is opposite to that of the
proton, with the difference accommodated by the orbital angular momentum.  
Such terms are responsible for reproducing the experimental feature
that $QF_2/F_1$ is approximately constant for  $Q^2\ge$ 
2 (GeV/c)$^2$\cite{Frank:1995pv,
Miller:2002qb,Miller:2002ig}.

The spin-independent momentum-space  wave function 
is  a function of the mass operator  $M_0$
 of a non-interacting system
of any  $P^\mu$:
$
M_0^2=
{(K_\perp^2+m^2\eta)/\eta(1-\eta)}+
{(k_\perp^2+m^2)/( \eta\xi(1-\xi))}, $
where $m$ is the $u,d$  quark mass.
 We take the 
 $S$-state orbital function $\Phi(M_0)$ to be of a  power law form:
$
\Phi(M_0)={N/(M^2_0+\beta^2)^{\gamma}}\;
$
that depends on $\beta,\gamma$ and the constituent quark mass $m$.
Parameters of the Light Front Cloudy Bag Model\cite{Miller:2002ig},
which  makes the Cloudy Bag Model\cite{cbm} relativistic, 
 are displayed in Table I. The effects of the pion cloud are unimportant at
high momentum transfer and are ignored here.

\begin{table}
  \centering
  \caption{Different parameter sets, $m,\beta$ in fm$^{-1}$. From
Ref.~\cite{Miller:2002ig}  }
  \vspace{0.1cm}
  \begin{tabular}{|l|rrr|}\hline
 {\em Set(legend)} & 
 $m$ &  $\beta$ & $\gamma$ \\ 
\hline
      1 dash & 1.8 &3.65  &4.1\\ 
      2 dot-dash & 1.8 &3.65&3.9 \\
      3 solid &1.7&  2.65 & 3.7\\ 
      \hline
      \end{tabular}
      \end{table}

Using 
light front dynamics  enables one to
relate the proton wave functions in different reference frames 
with a kinematic boost. 
 If the proton acquires a transverse
momentum 
by the  absorption of a quantity  of momentum,
$\bfDelta
\equiv(0,\bfDelta_{\perp})=\bfq-\bfq'$   by a quark,      
the effects of the boost are obtained 
by replacing the momenta $\bfk_\perp, \bfK_\perp$ 
by
$
\bfk_\perp,\; 
\bfK_\perp-\eta\bfDelta_{\perp}.$

The evaluation of the Compton scattering amplitude 
${\cal M}_{S',S}(\beps',\beps) $ is made with  
an impulse approximation in which the Compton scattering occurs 
via the
Born and crossed Born graphs of Fig.~1.
 The  antisymmetric nature of $\Psi$  allows us to take the 
scattering to occur on the third quark of charge $Q_3$. 
The incident (outgoing) photon
has four momentum and polarization vector $q, {\beps}$ ($q, \beps'$),
evaluated using the $\gamma p$ center of mass frame.
${\cal M}_{S',S}(\beps',\beps) $
depends on the initial spin $S$ and final spin $S'$ of 
the proton, as well as on $\beps,\beps'$:
\bea
{\cal M}_{S',S}(\beps',\beps)=
3Q_3^2 \int  d\eta d^2K_\perp \sum_{s,s'}\rho_{S',s';S,s}(\eta,
\bfK',\bfk)\bar{u}(K',s'){\cal O}_{\beps',\beps}(K',K)u(K,s)
,\label{ia}\eea
where $
\rho(\eta,\bfK',\bfk)\equiv\int d\xi d^2k_\perp 
\Psi^\dagger_{S',s'}
(\xi,\bfk_\perp,\eta,\bfK_\perp')\Psi_{S,s}(\xi,\bfk_\perp,\eta,
\bfK_\perp),$ and 
in which the repeated indices($s,s')$ represent both
 spin and isospin (charge) quantum numbers of the struck third
quark. The quantities  $K,K'$ are    four vectors given by
$K^\mu=((1-\eta)M_0,-\bfK)$ and
 $\notK u(K,s)=m u(K,s),\;\notK' u(K',s)=m u(K',s)$.
Satisfying 
the latter two relations is necessary to maintain gauge invariance. 
The operator ${\cal O}_{\beps',\beps}(K',K)$  represents 
 Compton scattering on  a  quark:
\begin{equation} 
{\cal O}_{\beps',\beps}(K',K) = 
\notepsp
  \frac{{1\over2}(K+q+K'+q')\cdot\gamma+m}{(K+q)^2 -m^2} \
\noteps +
\noteps\,  
  \frac{{1\over2}(K'-q+K-q')\cdot\gamma+m}{(K'-q)^2 -m^2} \
\notepsp 
. \label{fd}
\end{equation}
The numerators appearing
in \eqn{fd}) are displayed in 
a form symmetric with respect to the initial and final
states. This is necessary to maintain the  time reversal invariance of
the resulting amplitudes.
Deriving \eqn{ia}) from the full four-dimensional formalism involves
integration over the minus  components of the wave function and neglecting
modifications of  the intermediate propagator of the struck quark caused by
spectator quarks. Note also that the total momentum
of the proton does not enter into the expression (\ref{ia}), so that our
results are independent of frame.

Each of  $s,s',\beps,\beps'$, has two possible values, so 
 there are  16 amplitudes.
Using parity conservation and time reversal invariance reduces the
number of independent amplitudes to six\cite{rollnik}.
We calculate the 16
 amplitudes 
and demonstrate explicitly that there really are only six independent ones.
It is not obvious that the impulse approximation used in obtaining \eqn{ia})
will yield only six amplitudes. Applying  parity invariance immediately
reduces the number of  independent amplitudes to eight.  But
the effects of time reversal invariance are more difficult to 
satisfy.
This is because the sum of the quark minus-momenta is not equal to the
minus-momentum of the proton. Thus
 conservation of four-momentum
occurs only at the hadronic level, but not in the  
$\gamma q$ scattering.  
However, numerical calculations
show that   a reasonably accurate approximation can be made
that leads to the respect of time reversal invariance. Examine \eqn{ia}) and
 shift the variable of 
integration according to $\bfK_\perp\to\bfK_\perp +\eta\bfDelta/2$.
Then note that if the momentum
transfer $\Delta$ is large compared to typical momenta appearing in the
wave function, one may ignore the component of $\bfK$  parallel
to $\bfDelta$ in
  evaluating the matrix element
$\bar{u}(K',s'){\cal O}(\beps',\beps)u(K,s)$.
Numerical work shows that using 
this approximation doesn't change the computed values of observables
by significant amounts, but does reduce
the number
of independent amplitudes to exactly six, and also maintains gauge invariance.

The 
 relevant experimental observables involve   
  photons and protons  of a definite helicity. 
The $\gamma p$-cm 
helicity amplitudes are determined 
by making a unitary transformation on the 
spin amplitudes ${\cal M}_{S',S}(\beps',\beps)$:
\bea  \Phi_{\mu'\lambda',\mu\lambda}=
\sum_{S',S}T^*(\bfp')_{S'\lambda'}
{\cal M}_{S',S}(\beps_{\mu}',\beps_\mu)T_{S,\lambda}(\bfp),\label{unit}\eea
where $\mu,\lambda$ represent the helicity of the initial photon
and initial proton, and
$T_{S,\lambda}(\bfp)\equiv\bar{u}(p,s)u_H(p,\lambda)/(2M_p)$
in which $u$ represents an ordinary Dirac spinor and $u_H$ represents 
a helicity spinor.
Then the
differential  cross
section is expressed as:  
\bea
\frac{d\sigma}{dt}&=& \frac{1}{64\pi(s-m^2)^2} \Sigma_{\mu,\mu',\lambda,\lambda'}
        |\Phi_{\mu',\lambda',\mu\lambda}|^2 .
\label{eq:cross}\eea

\begin{figure}\unitlength1cm
\begin{center}
        \leavevmode
        \epsfxsize=0.75\textwidth
\begin{picture}(10,8)(-11,1)
  \includegraphics{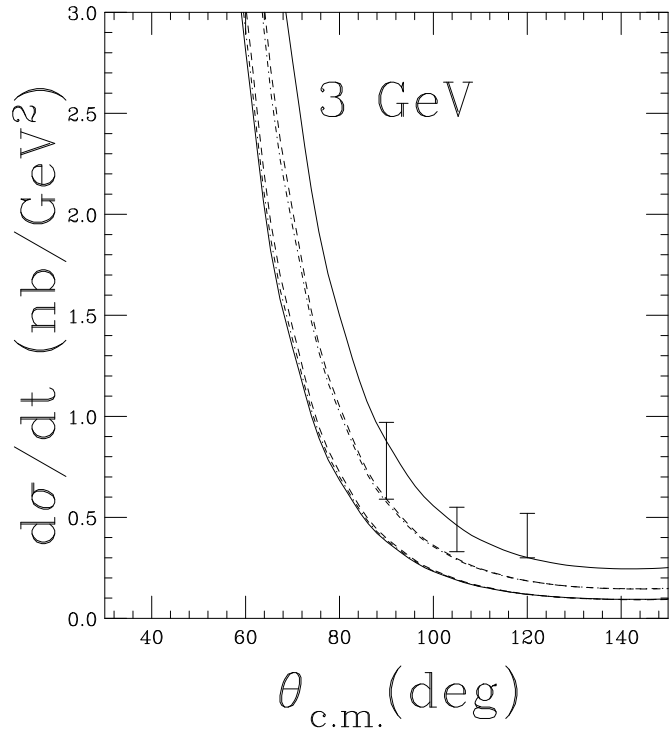}
\end{picture}
\end{center}
\caption{\label{fig:sigma3} Cross sections. 
The  data are 
from Ref.~\cite{CornellCS}.
Three curves are obtained with  parameters  defined in  Table 1 and 
three  are obtained
by increasing the quark mass $m$ by 10\%.  }
\end{figure}

\begin{figure}\unitlength1cm
\begin{center}
        \leavevmode
        \epsfxsize=0.75\textwidth
\begin{picture}(10,8)(-11,1)
  \includegraphics{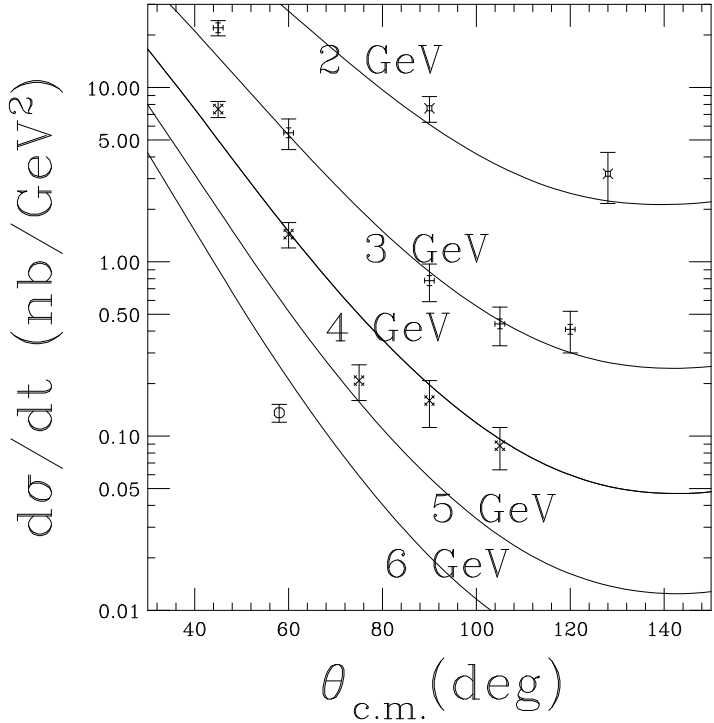}
\end{picture}
\end{center}
\caption{\label{fig:edep} Energy dependence of cross sections.
 The data are 
from Ref.~\cite{CornellCS}. The curves are  obtained
using set 3 of Table I, with an increase in the value of 
$m$ by 10\%.
}
\end{figure}

Our results for this are shown in Fig.~\ref{fig:sigma3}. Using the
parameter set of Table I leads to cross sections that are a bit too
small, 
but increasing the quark mass by 10\%, 
leads to qualitatively good agreement.
The dependence of $d\sigma/dt$ on photon laboratory energy 
in shown in  Fig.~\ref{fig:edep}.
There is a 20\% overall normalization uncertainty, and the data shown in
Figs.~\ref{fig:sigma3} and ~\ref{fig:edep} have been multiplied by
  0.8.

\begin{figure}\unitlength1cm
\begin{center}
        \leavevmode
        \epsfxsize=0.75\textwidth
\begin{picture}(10,8)(-11,1)
  \includegraphics{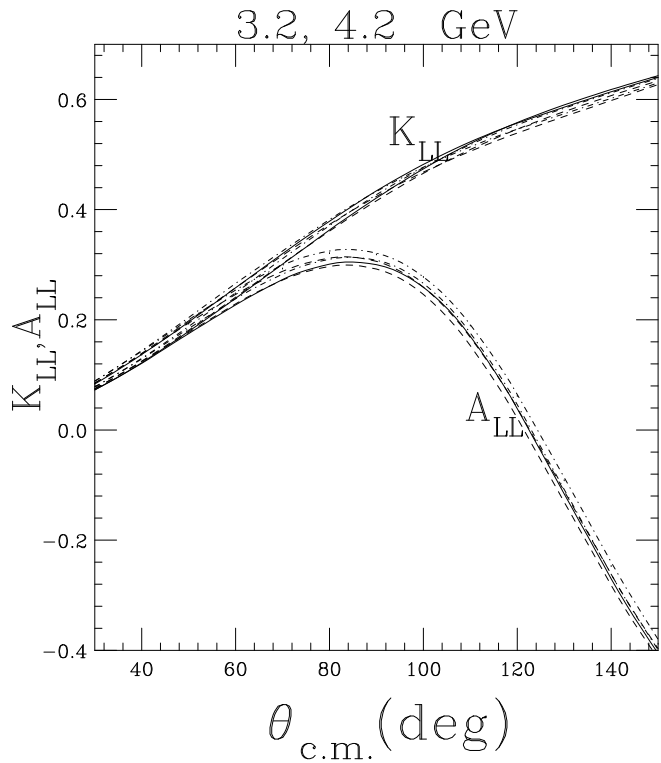}
\end{picture}
\end{center}
\caption{\label{fig:kae} $K_{LL},A_{LL}$ obtained using
the  parameter sets
of Fig.~\ref{fig:sigma3}. These are nearly independent of energy
and quark mass.
    }
\end{figure}

\begin{figure}\unitlength1cm
\begin{center}
        \leavevmode
        \epsfxsize=0.75\textwidth
\begin{picture}(10,8)(-11,1)
  \includegraphics{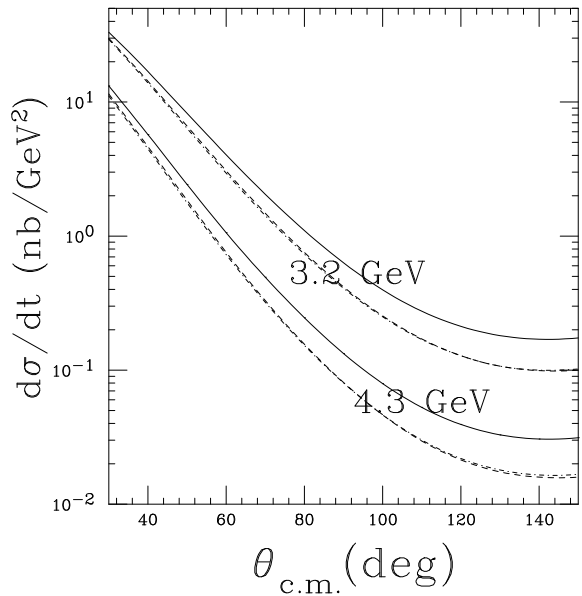}
\end{picture}
\end{center}
\caption{\label{fig:sigma3243} Computed values of ${d\sigma\over dt}$ the parameter sets
of Fig.~\ref{fig:sigma3} with an increased quark mass.
}
\end{figure}

The  spin dependent observables are defined according to 
Ref.~\cite{hkm}.
One set   involves 
 both
photon and proton helicities.
The  correlation  $K_{LL}$ between the helicities of the incoming
photon and the outgoing proton  is 
\bea
K_{LL}\frac{d\sigma}{dt}
      &=& \frac{d\sigma(\mu=+,\lambda'=+)}{dt}-
\frac{d\sigma(\mu=+,\lambda'=-)}{dt},
\eea
and 
 is especially interesting because the experimental result\cite{nathan} 
will be
announced soon.
The correlation, $A_{LL}$, between 
 the incident photon and the proton in the 
initial state is
\bea
A_{LL}\frac{d\sigma}{dt}
      &=&\frac{1}{2}\left[\frac{d\sigma(\mu=+,\lambda=+)}{dt}
-\frac{d\sigma(\mu=+,\lambda=-)}{dt}\right].
\eea
In the handbag approach of
Refs.\cite{Diehl,hkm}, the amplitudes
$\Phi_{--++}\equiv\Phi_2$ and $\Phi_{-++-}\equiv \Phi_6$ 
have the same magnitude, and this leads to
the prediction that $K_{LL}=A_{LL}$. 
Our results, for the  photon energies of immediate experimental interest
(3.2 and 4.3 GeV)\cite{nathan,bogdan},
 are displayed in Fig.~\ref{fig:kae}, and the related cross sections
can be found in Fig.~\ref{fig:sigma3243}. Examining Fig.~\ref{fig:kae}
shows that 
the predicted  values of $K_{LL}$ and $A_{LL}$
 do not depend on the quark mass, and have little variation with
energy. The values of   $K_{LL}$ are large and positive at large 
scattering angles. This is similar to the  $K_{LL}$
 of Refs.\cite{Diehl,hkm}, 
But their predicted  equality between  $K_{LL}$ and $A_{LL}$
does not hold here-- 
 for backward  scattering angles, we find $K_{LL}\approx-A_{LL}$.

This demands explanation. In Refs.~\cite{Diehl,hkm} the equality
between 
$K_{LL}$ and $A_{LL}$ arises from 
 neglecting  
the variation of the quark's minus and tranverse momenta, and 
 the effects of quark
orbital angular momentum that lead to vanishing   proton helicity flip
matrix elements.  
The  massless 
quarks 
are taken to move collinearly with the
proton which consequently does not change helicity.
 In our model, the role of orbital angular momentum
and  non-conservation of the proton helicity is the crucial
aspect in reproducing  the
proton form factors\cite{Miller:2002qb}.

Let's consider  scattering by 180$^\circ$ (back-angle scattering)
to illustrate
how it is that $K_{LL}\ne A_{LL}$. Examine \eqn{unit}). The transformation
matrices $T(\bfp)_{S\lambda}$ reduce to  overlaps between two-component 
spinors, in which $\boldsigma\cdot\widehat{\bfp}\vert\lambda\rangle=
\lambda\vert\lambda\rangle$. Then for back-angle scattering
$
T^*(\bfp')_{S'\lambda'}={1\over\sqrt{2}}(\delta_{S',+{1/2}}+\lambda'
\delta_{S',-{1/2}}),
T(\bfp)_{S\lambda}=
{1\over\sqrt{2}}(\delta_{S,+{1/2}}-\lambda
\delta_{S,-{1/2}}),
$
so that
$
\Phi_2={1\over2}\left[{\cal M}_{++}+{\cal M}_{--}
-{\cal M}_{+-}-{\cal M}_{-+}\right],\;
\Phi_6={1\over2}
\left[{\cal M}_{++}+{\cal M}_{--}+{\cal M}_{+-}+{\cal M}_{-+}\right].$
The dependence of ${\cal M}_{S'S}$ on the photon polarization
vectors is suppressed, as these are the same for $\Phi_{2,6}$.
Equality of $\Phi_2$ and $\Phi_6$ can only occur 
if  each of 
the proton spin flip matrix elements ${\cal M}_{+-},{\cal M}_{-+}$
vanish, or if  their sum vanishes. Inspection of \eqn{fd}) shows
that the spin-flip matrix elements do not vanish, and that
 the terms $\gamma\cdot(q+q')=\gamma^0 2 q^0$ lead to operators
$ (1+\sigma_x)$ evaluated between Pauli spinors. The operator
 $\sigma_x$ raises and
lowers spins with exactly strength, so that 
${\cal M}_{+-}+{\cal M}_{-+}$ does not vanish, $\Phi_2\ne\Phi_6$, and 
$A_{LL}\ne K_{LL}.$ 

Another way to understand this inequality is to
examine the numerical effect of reducing the quark mass towards
 0. We find that this  causes $A_{LL}$ to aproach $K_{LL}$.

\begin{figure}\unitlength1cm
\begin{center}
        \leavevmode
        \epsfxsize=0.75\textwidth
\begin{picture}(10,8)(-11,1)
  \includegraphics{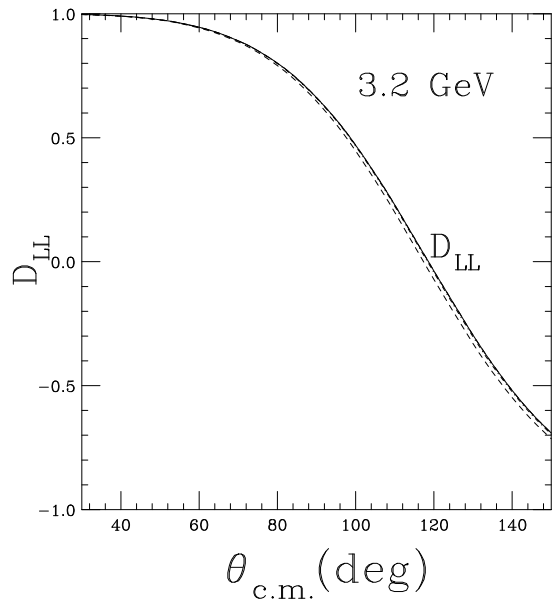}
\end{picture}
\end{center}
\caption{\label{fig:dp} Computed values of $D_{LL}$, using  parameter sets
of Table~I.}
\end{figure}

There are other polarization variables\cite{hkm}.
The helicity transfer from the
incoming to the outgoing photon is given by  
$D_{LL}
{d\sigma}/{dt} = 
\left({d\sigma(\mu=+,\mu'=+)}/{dt}
- 
{d\sigma(\mu=+,\mu'=-)}/{dt}
\right).$ 
Ref.~\cite{hkm} finds that
$      D_{LL} \approx  1. $
The polarization of the incoming proton is defined by
$ P
{d\sigma}/{dt}= 
\frac12 \left[ 
{d\sigma(\uparrow)}/{dt}
                          - 
{d\sigma(\downarrow)}/{dt} \right]
. $ 
In Ref.~\cite{hkm} small corrections 
lead to estimating that $P\approx 3\%$. Our result is that
$P=0$. Our prediction for $D_{LL}$,
shown in Fig~\ref{fig:dp}, is  that it has significant deviations
from unity
at large scattering
angles. 

Now consider sideways proton spin directions.
The correlation between the helicity of the incoming photon
and the sideways ($S$) polarization of the incoming proton, parallel
 or anti-parallel 
to the $S$-direction is 
defined\cite{hkm} as $A_{LS}$, and the one for
 the 
sideways polarization of the outgoing proton is $K_{LS}$.
We find $K_{LS}=0$ and  $A_{LS}=0$, and that
the incoming photon
asymmetry $\Sigma$ \cite{hkm} vanishes.

Let's summarize.
\poinc invariance,
gauge invariance, conservation of parity and time reversal invariance
are  respected in our  impulse approximation evaluation of the
handbag diagrams. Proton wave functions, previously constrained
by comparison with  measured form factors, that incorporate the influence
of quark orbital angular momentum (and the corresponding violation
of proton helicity conservation) are used.  Computed cross sections
are in reasonably good agreement with early measurements. 
The value of $K_{LL}$ is large and positive for scattering at large angles.
	In contrast with earlier work, we find that
 $K_{LL}\ne A_{LL}$, and $D_{LL}\ne1$ at large scattering angles.
With our  model functions,
 photon laboratory energies of 6 GeV or  less are too low
for the simplifying assumptions that lead to proton helicity conservation
to be valid. Future experiments that measure
$A_{LL}$ or $D_{LL}$ 
can determine whether or not proton helicity conservation holds in Compton
scattering at any given energy.

\begin{acknowledgments}
I thank B. Wojtsekhowski for suggesting this project, and P. Kroll for
useful discussions.
I am  grateful to the  ECT*, 
the 
  National Institute for Nuclear Theory at the 
UW, the NSTG at 
LBL 
 and the CSSSM 
for providing hospitality during the course of  this
work. The Guggenheim foundation, on the other hand, was of no help at all.
This work is partially  supported by 
the USDOE. 
\end{acknowledgments}

\bibliographystyle{unsrt}

\end{document}